\documentclass[12pt,a4paper]{article}
\usepackage{mathrsfs}
\usepackage{epsfig}
\begin{document}

\title{ Light axigluon and single top production at the $LHC$ }
\author{ Chong-Xing Yue, Shi-Yue Cao, Qing-Guo Zeng\\
{\small Department of Physics, Liaoning  Normal University, Dalian
116029, P. R. China}
\thanks{E-mail:cxyue@lnnu.edu.cn}}
\date{\today}

\maketitle
\begin{abstract}
The light axigluon model can explain the Tevatron $t\overline{t}$ forward-backward asymmetry
and at the same time satisfy the constraints from the electroweak precision measurement and
the $ATLAS$ and $CMS$ data, which induces the flavor changing ($FC$) couplings of axigluon
with the $SM$ and new quarks. We investigate the effects of these $FC$ couplings on the s- and t-channel
single top productions at the $LHC$ and the $FC$ decays $Z\rightarrow \overline{b}s+b\overline{s}$,
$t\rightarrow c\gamma$ and $cg$. Our numerical results show that the light axigluon can
give significantly contributions to single top production and the rare top decays $t\rightarrow c\gamma$ and $cg$.

\vspace{01.0cm}
{\bf Key words:} light axigluon, $FC$ couplings, single top production, rare top decays

\vspace{0.2cm} \noindent
 {\bf PACS numbers}: 12.90.+b, 14.70.Pw, 14.65.Ha

\end{abstract}
\newpage
\noindent{\bf 1. Introduction }\vspace{0.5cm}

The standard model $(SM)$ of particle physics has been proven to be
extremely successful describing collider experimented data so far.
Even the discovery of a Higgs-like particle [1, 2] has confirmed the
validity of the $SM$ at the Fermi scale. However, the $SM$ suffers
from a key theoretical drawback, the so-called "hierarchy" problem, which
means that it could be a low-energy effective theory valid only up to
some cut-off energy scale $\Lambda $, about $TeV$ scale. So new physics
beyond the $SM$ would be in an energy range accessible at the $LHC$ and
might be discovered in coming years, although, at the moment, there is
not any collider hint of new physics at the $LHC$.

There are various new physics models extending the gauge group of the strong
interaction sector give rise to massive color-octet vector boson, for
example, the topcolor models [3] and chiral color models [4]. Other examples
include the extra dimensional models [5] and technicolor [6], which predict
the existence of the Kaluza-Klein (KK) gluons and technirhos, respectively.
Among these color-octet vector bosons, the new paricles with axial-vector
couplings to the $SM$ quarks are called "axigluons", which might explain
the anomalous forward-backward asymmetry ($FBA$) in the $t\overline{t}$
production observed at the Tevatron [7]. So far, there has been a significant
amount of works to explain the $t\overline{t}$ $FBA$ via axigluons, for example
see [8, 9, 10, 11, 12, 13]. Furthermore, the light axigluon $A$ with a mass $M_{A}$ in the
range from $100GeV$ to $400GeV$ can explain the $t\overline{t}$ $FBA$ and satisfy
the constraints from the $ATLAS$ and $CMS$ data [14, 15], as long as its decay width
is large and its couplings to the $SM$ quarks are relatively small [9, 10, 11, 12].

Top quark physics is expected to be a window to any new physics beyond the electroweak
scale. At $LHC$ energies, top quark is copiously produced both in pair and single
productions, which allows for an unprecedented precision in the study of top observables,
such as its couplings and rare decays [16]. At hadron colliders, single top quark
production is an important process in probing the mechanism of electroweak symmetry
breaking ($EWSB$), providing informations  complementary to those that can be obtained
from top pair production [17]. Single top production is also very sensitive to new physics
effects, whose strength can be assessed by precise measurement of the production cross section.

Single top production at hadron colliders has been observed in three channels: s-channel, t-channel
[18, 19] and $tW$ associated production channel [20], which accord with the $SM$ predictions within
experimental uncertainties. $ATLAS$ and $CMS$ collaborations have started searching for the new physics
effects on single top production.

Inspired by the solution of the light axigluon to the $t\overline{t}$ $FBA$,  some axigluon-mediated
phenomena are studied in this paper. We consider the contributions of the light axigluon with flavor changing ($FC$) couplings to the
$SM$ and new quarks to the $FC$ decays $Z\rightarrow \overline{b}s (b\overline{s})$, the s- and t-channel single
top productions, and rare top decays $t\rightarrow c\gamma$ and $cg$ in the context of the light axigluon model
proposed by Tavares and Schmaltz [10]. The constraints on this new physics model from the electroweak precision
observables and the relevant data given by hadron colliders are taken into account in our numerical calculations.

The rest of this paper is organized as follows: After reviewing the basic ingredients of the light axigluon model,
in section 2, we calculate the contributions of the light axigluon to the $FC$ decays $Z\rightarrow \overline{b}s$ and $ b\overline{s}$. Corrections of the light axigluon to the cross sections of the s- and t-channel single top productions
at the $LHC$ are studied in section 3. The branching ratios of the rare top decays $t\rightarrow c\gamma$ and $cg$
induced by light axigluon exchange are given in section 4. Section 5 is devoted to simple summary.

\vspace{0.5cm} \noindent{\bf 2. Light axigluon  and the $FC$ decays $Z\rightarrow \overline{b}s$ and $ b\overline{s}$  }

\vspace{0.5cm}The light axigluon model [10] is based on the gauge group $G=SU(3)_{1}\times SU(3)_{2}\times SU(2)\times U(1)_{Y} $,
where $SU(2)\times U(1)_{Y}$ is the conventional electroweak group and the extended gauge group $SU(3)_{1}\times SU(3)_{2}$ is
spontaneously broken to the $QCD$ gauge group $SU(3)_{C}$ by the vacuum expectation value ($VEV$) of a bifundamental scalar $\phi$.
This breaking pattern yields two mass eigenstates of color-octet gauge bosons. One is massless particle, which can be identified
with the $SM$ gluon, and the other is massive particle, which is called the light axigluon $A$. For its couplings to the $SM$ quarks,
there are the vector coupling $g_{V}\approx0$ and the axial-vector coupling $g_{A}\neq 0$ in the case of assuming approximately parity
symmetry. In order to cancel the gauge anomaly, the extra up- and down-type quarks are introduced into this model, and the
lepton sector is exactly same as that of the $SM$. To explain the $t\overline{t}$ $FBA$, the axigluon $A$ should have mass below
$450GeV$, while should be broad with $\Gamma_{A}/M_{A}\sim10\sim20\%$, where $\Gamma_{A}$ and $M_{A}$ represent its total decay width and
mass, respectively.

In the original light axigluon model [10], the authors assume the existence of an exact global symmetry of the
axigluon couplings, and thus the light axigluon only has flavor universal couplings to the $SM$ quarks. In fact,
this global symmetry is only approximate and there is mixing between new and ordinary  quarks, which can induce
flavor changing neutral currents ($FCNCs$) at tree level [21]. The new and ordinary quarks have same $SU(2)\times U(1)$
charge, their mixing does not give rise to the $FC$ $Z$ couplings at tree level. The new scalars can not induce $FCNCs$,
thus the non-universal axigluon couplings are the main source of $FCNC$ for this model.

In this paper we will not assume the existence of an exact global symmetry of the axigluon couplings, which allows $FC$ couplings of
the axigluons to the $SM$ quarks. If one assumes that these $FC$ couplings  are only axial-vector couplings, which
are similar with their  flavor conserving couplings to the $SM$ quarks, then the axial-vector couplings of the light
axigluon to the $SM$ quarks can be general given by the Lagrangian
\begin{eqnarray}
{{\cal
L}\supset g_{s}[\overline{u}_{i}\gamma_{\mu}\gamma_{5}(g^{u_{i}}_{A}\delta_{ij}+\varepsilon^{ij}_{u})u_{j}A^{\mu}+\overline{d}_{i}\gamma_{\mu}\gamma_{5}((g^{d_{i}}_{A}
\delta_{ij}+\varepsilon^{ij}_{d})d_{j}A^{\mu}]},
\end{eqnarray}
where $A^{\mu}$ is the light axigluon, $g_{s}$ is the $QCD$ coupling constant, $u_{i}$ and $d_{i}$ are the
$SM$ up- and down-type quarks, respectively. In above equation, we have neglected the color and spinor indices.
$g^{u_{i}}_{A}$ and $g^{d_{i}}_{A}$ are the flavor independent coupling constants and there are $g^{u_{i}}_{A}=g^{d_{i}}_{A}= g^{q}_{A}$ [10]. The $FC$ coupling constants $\varepsilon^{ij}_{u}$ and $\varepsilon^{ij}_{d}$, which arise from flavor symmetry breaking of new and light quarks,
are given by the matrices
\begin{eqnarray}
\varepsilon_{u}=
\left( \begin{array}{ccc}0&g^{uc}
 &g^{ut}\\(g^{uc})^{\ast}&0&g^{ct}\\(g^{ut})^{\ast}&(g^{ct})^{\ast}&0\end{array}\right) , ~~
\varepsilon_{d}=
\left( \begin{array}{ccc}0&g^{ds}
 &g^{db}\\(g^{ds})^{\ast}&0&g^{bs}\\(g^{db})^{\ast}&(g^{bs})^{\ast}&0\end{array}\right).
\end{eqnarray}

The couplings of the axigluon to a pair of ordinary quarks and to the corresponding partners have opposite sign. So, in order to get suppressed couplings of the ordinary quarks to the axigluon, the extra quarks and the $SM$ quarks should have mixing [10, 12, 22]. The mixing can be  obtained by adding a Yukawa coupling involving a scalar field $\phi$ in addition to the quark field of $Q'$ with $Q$. After the spontaneous breakdown of $SU(3)_{1}\times SU(3)_{2}\rightarrow SU(3)_{C}$ induced by the $VEV$ for $\phi$, the new quarks from the line combinations of $Q'$ and $Q$ get masses, while their orthogonal combinations correspond to the $SM$ quarks remain massless, which get
masses from the $SM$ Higgs $VEV$ via Yukawa couplings. In the mass eigenstates, the mixing couplings of  the axigluon to ordinary and  new quarks, which are assumed to be axial-vector couplings, can be general written as
\begin{eqnarray}
{{\cal
L^{' }}\supset g_{s}g^{mix}_{A}[\overline{U}_{Hi}\gamma_{\mu}\gamma_{5}(\varepsilon^{ij}_{Hu})u_{j}A^{\mu}+\overline{D}_{Hi}\gamma_{\mu}\gamma_{5}(\varepsilon^{ij}_{Hd})d_{j}A^{\mu}]}.
\end{eqnarray}
$U_{Hi}$ and $D_{Hi}$ represent the up-type and down-type new quarks, respectively. For the mixing coupling constant $ g^{mix}_{A} $, there is the relation  $ (g^{mix}_{A})^{2} + (g^{q}_{A})^{2}= 1 $. For the two matrices $\varepsilon_{Hu}$ and $\varepsilon_{Hd}$, they are related through the $SM$ $CKM$ matrix:  $\varepsilon^{+}_{Hu}\varepsilon_{Hd}= V_{CKM}$, which is similar with the case for the mixing between the T-odd and T-even quarks in the $LHT$ model [23]. In this paper,  we assume that both
$\varepsilon_{Hu}$ and $\varepsilon_{Hd}$ are nearly equal to the identity matrix, which provides us with a set of minimal flavor mixing scenarios. We take as examples two simple cases:

Case I $\varepsilon_{Hu} = I, ~~ \varepsilon_{Hd}= V_{CKM}$,

Case II $\varepsilon_{Hd} = I, ~~ \varepsilon_{Hu}= V_{CKM}$.

In case I, the mixing coupling $ g^{Qq}_{A} $ has no contributions to $D^{0}-\overline{D^{0}}$ mixing, while contributes to $B_{q}^{0}-\overline{B_{q}^{0}}$ and  $K^{0}-\overline{K^{0}}$ mixings. For case II,  it is obvious that the mixing coupling $ g^{Qq}_{A} $ can only contribute to $D^{0}-\overline{D^{0}}$ mixing.
Reference [21] has obtained the constraints on the mixing matrix $\varepsilon_{d}$ by using
the available data from neutral meson mixings, such as $B_{q}^{0}-\overline{B_{q}^{0}}$, $K^{0}-\overline{K^{0}}$
and $D^{0}-\overline{D^{0}}$ mixings. Taking into account of these constants, in this section,
we calculate the branching ratios of the $FC$ decays $Z\rightarrow \overline{b}s$ and $ b\overline{s}$ given by axigluon
exchange as shown in Fig.1. The self-energy diagrams Fig.1(b) and (c) contribute a finite field renormalization
and the individual diagrams are finite [24]. To fulfill the broad width of the axigluon, the first and second generation new quarks should be degenerate and
lighter than the axigluon, while the third  generation new quarks must be heavier [10]. So we think that the contributions of the third  generation new quarks to the $FC$ decays $Z\rightarrow \overline{b}s (b\overline{s})$ decouple and only consider the contributions of the first and second generation new quarks. In our numerical estimation, we will take $M_{D_{H1}}= M_{D_{H2}}= M_{H} =0.2 M_{A}$. In this case, one can safely neglect the phase space suppression effect for the axigluon decaying to one new quark and one ordinary quark and there should be $\Gamma_{A}/M_{A}\sim10\sim20\%$.

\vspace{1cm}
\begin{figure}[htb]
\vspace{-0.5cm}
\begin{center}
 \epsfig{file=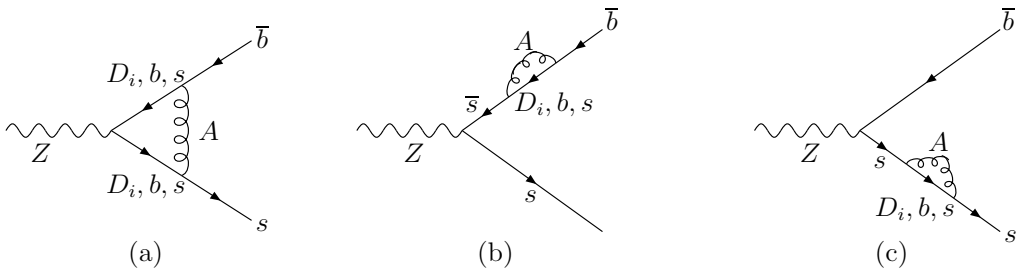,width=380pt,height=100pt}
\vspace{-0.5cm} \caption{  One-loop Feynman diagrams for the $FC$ decay $Z\rightarrow \overline{b}s$ induced by light axigluon \hspace*{1.7cm} exchange. } \label{ee}
\end{center}
\end{figure}
\vspace{-0.5cm}

The light axigluon model  predicts the existence of new scalar, which also has the mixing couplings to new and ordinary quarks. However, it can not induce $FC$ couplings at tree level and thus in this paper we neglect the effects of the new scalar on the $FC$ processes $Z\rightarrow \overline{b}s$ and $ b\overline{s}$.

The corrections of color-octet gauge boson to the $Zb\overline{b}$ coupling are firstly studied by Ref.[25] in the
context of topcolor models, which contain only the leading-logarithmic contributions. The full one-loop results for
the corrections of the axigluon to the $Zb\overline{b}$ coupling are given in Refs.[11, 12] in the case of neglecting
the bottom quark mass. Ref.[12] have further computed the contributions from new quarks and new scalar to the $Zb\overline{b}$ coupling and find that the two kinds of contributions have opposite sign and the effect of new scalar is much smaller than that of new quarks. Following Refs.[11, 12], we can straightforwardly calculate the contributions of the light axigluon model to the $FC$ couplings $Z\overline{b}s$ and $Zb\overline{s}$. Then, the effective $Z\overline{b}s$ coupling
can be written as
\begin{eqnarray}
g_{P}^{Zbs}=\frac{\alpha_{s}}{3\pi}g_{P}^{Zbb}[2g_{P}^{Abb}g_{P}^{Abs}\kappa(x_{z})+ (g^{mix}_{A})^{2}\kappa(x_{z}, x_{h})(\varepsilon^{*13}_{Hd}\varepsilon^{12}_{Hd}+\varepsilon^{*23}_{Hd}\varepsilon^{22}_{Hd})],
\end{eqnarray}
where $P=L$ and $R$.  $g_{P}^{Zbb}$ and $g_{P}^{Abb}$ represent the couplings of the
gauge boson $Z$ and axigluon $A$ to the bottom quark pairs, respectively. The explicit
expressions of the factors $\kappa(x_{Z})$ and $\kappa(x_{z}, x_{h})$ have been given in Ref.[12]. Since the couplings of the axigluon to pair of ordinary quarks and pair of new quarks are flavor universal and the new and ordinary quarks have same $SU(2)\times U(1)$
charge, in above equation we have added the contributions of the ordinary quarks $b$ and $s$, and taken
\begin{eqnarray}
{g_{L}^{Zbb}=g_{L}^{ZD_{i}D_{i}}=\frac{e}{4S_{W}C_{W}}(1-\frac{2}{3}S_{W}^{2})}, ~~  {g_{R}^{Zbb}=g_{R}^{ZD_{i}D_{i}}=-\frac{e}{4S_{W}C_{W}}\cdot\frac{2}{3}S_{W}^{2}},
\end{eqnarray}
where $i=1 $ and $2$, $S_{W}=\sin \theta_{W}$ and $C_{W}=\cos \theta_{W}$, $\theta_{W}$ is the Weinberg angle. The $FC$ coupling $g_{P}^{Abs}$ can contribute to $B_{s}^{0}-\overline{B_{s}^{0}}$ mixing at tree level and its upper bound has been obtained by Ref.[21] as  $|g_{L}^{bs}|=|g_{R}^{bs}|=|g_{A}^{bs}|\leq 1.83\times 10^{-3}$. In fact, for the case I, the new quarks can also generate contributions to $B_{s}^{0}-\overline{B_{s}^{0}}$
mixing via box diagrams that contain the light axigluon  and new quark. However, the contributions from box diagrams are suppressed with respect to axigluon tree-level contributions by a loop factor $1/(16\pi^{2})$ and two additional mixing matrix elements $\varepsilon^{i3}_{Hd}$ and $\varepsilon^{i2}_{Hd}$. Therefore they cannot compete with the latter and are negligible. As numerical estimation, we will take $g_{A}^{bs}=1.83\times 10^{-3}$, $g_{L}^{Abb}=-g_{R}^{Abb}=g_{A}^{q}$.

In the $SM$, the $FC$ decay $Z\rightarrow \overline{b}s+b\overline{s}$ originates
from one loop diagrams with branching ratio $\sim 3\times10^{-8}$ [26]. For future
linear collider ($ILC$), the expected sensitivity to the branching ratios of rare $Z$
decays can be improved from $10^{-5}$ at the $LEP$ to $10^{-8}$ at the Giga $Z$ [27].
The new physics effects might be detectable via $Z\rightarrow bs$ if it indeed affects
this decay. A lot of theoretical studies involving the $FC$ decay $Z\rightarrow bs$
have been given within some popular models beyond the $SM$, where its branching ratio can be
significantly enhanced [28].
\begin{figure}[htb]
\begin{center}
 \epsfig{file=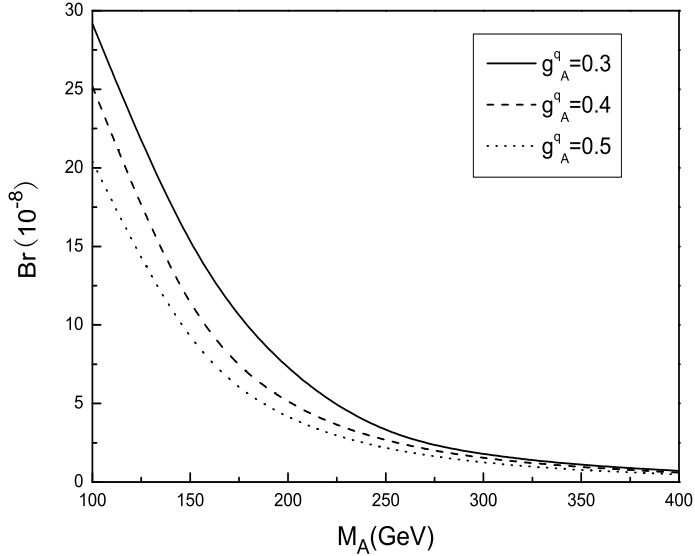,width=300pt,height=260pt}
\vspace{-0.5cm} \caption{Variation of the branching ratio $Br(Z\rightarrow \overline{b}s+b\overline{s})$ with the axigluon mass $M_{A}$ \hspace*{1.7cm} for $g_{A}^{bs}=1.83\times10^{-3}$, $\varepsilon_{Hd}= V_{CKM}$ and three values of the coupling parameter  \hspace*{1.7cm} $g_{A}^{q}$. } \label{ee}
\end{center}
\end{figure}

Using the effective couplings $g_{L}^{Zbs}$ and $g_{R}^{Zbs}$ given by Eq.(4), we can easily
calculate the partial width $\Gamma(Z\rightarrow \overline{b}s+b\overline{s})$. The numerical
results for the branching ratio $Br(Z\rightarrow \overline{b}s+b\overline{s})=\Gamma(Z\rightarrow \overline{b}s+b\overline{s})/\Gamma_{total}$ are shown in Fig.2, in which we have taken the $SM$
input parameters as: $\alpha_{s}(m_{Z})=0.118$, $S_{W}^{2}=0.231$, $\Gamma_{total}=2.4945GeV$, and
$M_{Z}=91.1875GeV$ [29]. If the light axigluon can explain the $t\overline{t}$ $FBA$ and at the same
time satisfy the constraints from the electroweak precision observables and the relevant data given by
hadron colliders, its mass should be in the range of $100GeV\sim400GeV$, its total decay width $\Gamma_{t}^{A}=(0.1\sim0.2)M_{A}$
and the flavor conserving coupling $g_{A}^{q}$ might be in the range of $0.3\sim0.5$ [9, 10, 11, 12]. In
our numerical estimation we have considered the effects of the axigluon width and taken $\Gamma_{t}^{A}=0.1 M_{A}$. For  the mixing between the $SM$ and new quarks, we have taken case I and assumed $M_{H}=0.2M_{A}$. One can
see from Fig.2 that, in most of the parameter space, the value of the branching ratio $Br(Z\rightarrow \overline{b}s+b\overline{s})$
is smaller than $1\times10^{-8}$, which is still below the $SM$ prediction. So considering
the constraints of $B_{s}^{0}-\overline{B_{s}^{0}}$ mixing on the $FC$ coupling $g_{A}^{bs}$, the contribution of the
light axigluon to the rare decays $Z\rightarrow \overline{b}s$ and $ b\overline{s}$ is very difficult to be detected in near future. Certainly, if we assume $\varepsilon_{Hd}\neq V_{CKM}$, the numerical results should has some changes.

\vspace{0.5cm} \noindent{\bf 3. The $FC$ couplings of the light axigluon $A$ and single top production at the  \hspace*{0.5cm}$LHC$ }

\begin{figure}[htb]
\begin{center}
 \epsfig{file=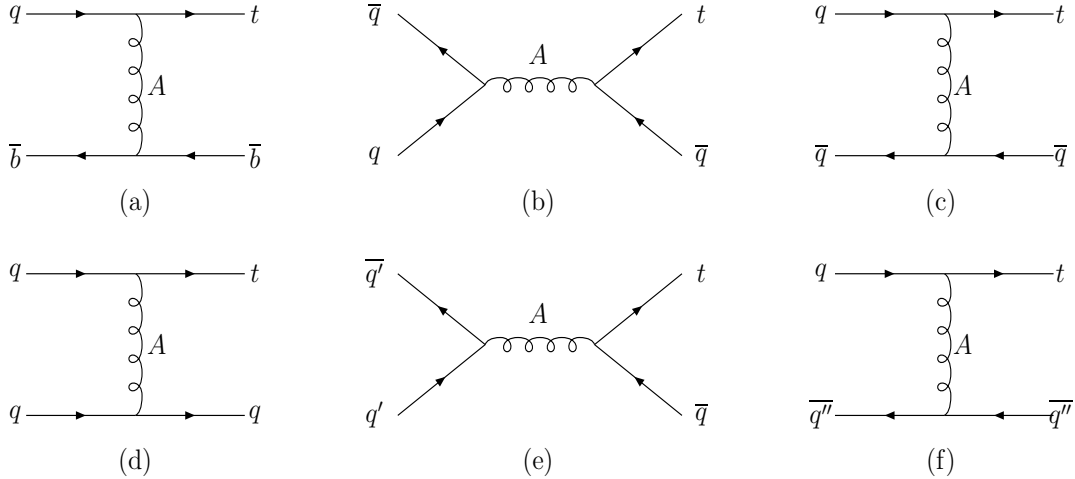,width=400pt,height=180pt}
\vspace{-0.5cm} \caption{Leading order Feynman diagrams for $t\overline{b}$ and $t\overline{j}$ production contributed by the \hspace*{1.7cm} $FC$ couplings $g_{A}^{tq}$, in which $q= u, c$, $q'=d, s, b$, and $q''=d, s$. } \label{ee}
\end{center}
\end{figure}

\vspace{0.5cm} In the $SM$, single top production dominantly occurs through electroweak processes, which are customary
divided into three production channels: t-channel exchange of a space-like W boson, s-channel production and
decay of a time-like W boson, and associated production of a top quark and an on-shell W boson. These partonic
processes have their own distinct kinematics and do not interfcere with each other. Both at Tevatron and the $LHC$,
the t-channel process is dominant one, which in five flavor ($5F$) scheme proceeds via the partonic processes
$qb\rightarrow q't$ and $\overline{q}b\rightarrow \overline{q'}t$ for single top production, and
$q\overline{b}\rightarrow q'\overline{t}$ and $\overline{q}$$\overline{b}$ $\rightarrow $ $\overline{q'}$ $ \overline{t}$ for
single antitop production. The s-channel partonic processes are $q\overline{q'}\rightarrow t\overline{b}$ and
$\overline{q}q'\rightarrow \overline{t}b$ for single top and antitop productions, respectively. The contributions
of charged and neutral color-octet vector bosons to top pairs and single top production has been studied in
Refs.[13, 30]. In this section we will consider the corrections of the light axigluon to the s- and t-channel single
top productions via the $FC$ couplings $g_{A}^{tq}$ with $q= u$ or $c$. The relevant Feynman diagrams are shown in Fig.3.

For the partonic process $q\overline{b}\rightarrow t\overline{b}$ as shown in Fig.3 (a), the differential cross
section with respect to emerging angle of the single top quark $\cos\theta_{t}$ can be written as
\begin{eqnarray}
{\frac{d\sigma(t\overline{b})}{d\cos\theta_{t}}=\frac{2\pi\alpha_{s}^{2}\beta(g_{A}^{tq})^{2}(g_{A}^{b})^{2}}{9\hat{s}}
P_{t}[\hat{s}(\hat{s}-m_{t}^{2})+\hat{t}(\hat{t}-m_{t}^{2})]}.
\end{eqnarray}
The partonic process $q\overline{q}\rightarrow t\overline{q}$ is composed of the s- and t-channel diagrams
corresponding to Fig.3 (b) and 3 (c). Its differential cross section is given by
\begin{eqnarray}
\frac{d\sigma(t\overline{q})}{d\cos\theta_{t}}=\frac{2\pi\alpha_{s}^{2}\beta(g_{A}^{tq})^{2}(g_{A}^{q})^{2}}{9\hat{s}}
\{P_{s}[\hat{u}(\hat{u}-m_{t}^{2})+\hat{t}(\hat{t}-m_{t}^{2})]
\nonumber\\-\frac{P_{s}P_{t}}{3}(\hat{s}-M_{A}^{2})(\hat{t}-M_{A}^{2})\hat{u}(\hat{u}-m_{t}^{2})\hspace*{1.5cm}
\nonumber\\+P_{t}[\hat{s}(\hat{s}-m_{t}^{2})+\hat{u}(\hat{u}-m_{t}^{2})]\}.\hspace*{2.5cm}
\end{eqnarray}
The differential cross section of the t+u channel partonic process $qq\rightarrow t+q$ can be written as
\begin{eqnarray}
\frac{d\sigma(tq)}{d\cos\theta_{t}}=\frac{2\pi\alpha_{s}^{2}\beta(g_{A}^{tq})^{2}(g_{A}^{q})^{2}}{9\hat{s}}
\{P_{t}[\hat{u}(\hat{u}-m_{t}^{2})+\hat{s}(\hat{s}-m_{t}^{2})]
\nonumber\\+P_{t}P_{u}(\hat{t}-M_{A}^{2})(\hat{u}-M_{A}^{2})\hat{s}(\hat{s}-m_{t}^{2})\hspace*{1.6cm}
\nonumber\\+P_{u}[\hat{t}(\hat{t}-m_{t}^{2})+\hat{s}(\hat{s}-m_{t}^{2})]\}.\hspace*{2.7cm}
\end{eqnarray}
The differential cross section for the s-channel partonic $\overline{q'}q'\rightarrow t\overline{q}$ as shown in Fig.3 (e) is
given by
\begin{eqnarray}
\frac{d\sigma_{s}(t\overline{q})}{d\cos\theta_{t}}=\frac{2\pi\alpha_{s}^{2}\beta(g_{A}^{tq})^{2}(g_{A}^{q'})^{2}}{9\hat{s}}
P_{s}[\hat{u}(\hat{u}-m_{t}^{2})+\hat{t}(\hat{t}-m_{t}^{2})].
\end{eqnarray}
The explicit expression of the differential cross section for the t-channel $q\overline{q''}\rightarrow t\overline{q''}$
is same as that for the process $q\overline{b}\rightarrow t\overline{b}$, as long as replace the initial state $b$ quark by
the quark $q''$ ($d$ or $s$). In above equations, $\beta=1-\frac{m_{t}^{2}}{\hat{s}}$, $\hat{s}$, $\hat{t}$, and $\hat{u}$
are the usual Mandelstam variables,
\begin{eqnarray}
{P_{i}=\frac{1}{(i-M_{A}^{2})^{2}+M_{A}^{2}\Gamma_{A}^{2}}}\hspace*{1.0cm}with\hspace*{0.4cm} i=\hat{s},\hspace*{0.1cm} \hat{t},or \hspace*{0.1cm}\hat{u}.
\end{eqnarray}

Using above equations we can calculate the cross sections of $tb$ and $tj$ production at the $LHC$ induced by
the light axigluon with  the $FC$ coupling $g_{A}^{tq}$. In our numerical calculations, we use the leading order parton
distribution function of CTEQ6L1 [31] and choose the factorization and renormalization scales to be $\mu_{f}=\mu_{r}=m_{t}/2$
with $m_{t}=173GeV$. Our numerical results are added $t\overline{b}$ and $\overline{t}b$ for the process $pp\rightarrow tb$,
and similar for $tj$ production with $j=u, c, d,$ and $s$. It is obvious that the production cross sections depend on the
mass parameter $M_{A}$, the coupling parameters $g_{A}^{tq}$ and $g_{A}^{q}$, where we have taken $g_{A}^{tu}=g_{A}^{tc}$
and the flavor conserving coupling $g_{A}^{q}$ being flavor universal.
\begin{figure}[htb]
\begin{center}
 \epsfig{file=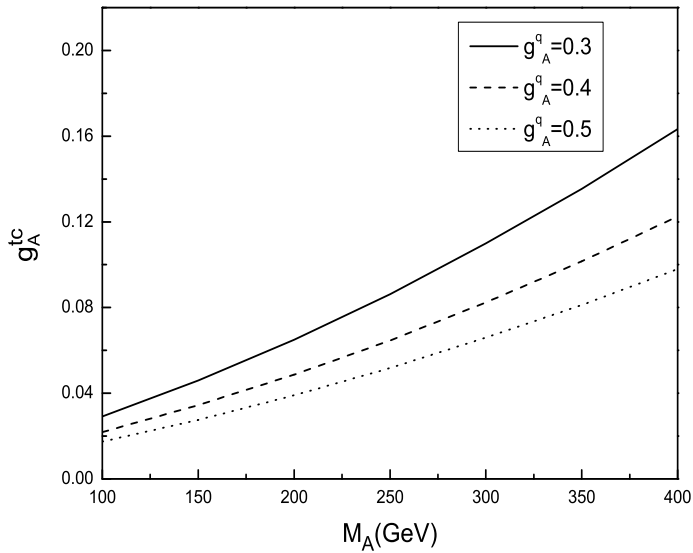,width=300pt,height=260pt}
\vspace{-0.5cm} \caption{In the case of $\delta\sigma^{s}/\sigma_{SM}^{s}=10\%$, the $FC$ coupling $g_{A}^{tq}$ as function of the axigluon \hspace*{1.7cm} mass $M_{A}$ for $g_{A}^{q}=0.3$(solid line), $0.4$(dashed line) and $0.5$(dotted line). }
\label{ee}
\end{center}
\end{figure}

In the $SM$, single top production at hadron colliders was first considered in Ref.[32]. Now the production cross sections
for the s- and t-channels have been calculated up to next-to-next-to leading logarithm ($NNLL$) accuracy [33]:
$\sigma_{s}=1.04\pm 4\%$ $pb$ and $\sigma_{t}=2.26\pm 5\%$ $pb$ at Tevatron with the centre-of-mass ($c. m.$) energy
$\sqrt{s}=1.96TeV$ and $\sigma_{s}=12\pm 6\%$ $pb$ and $\sigma_{t}=243\pm 4\%$ $pb$ at the $LHC$ with $\sqrt{s}=14 TeV$.
The s- and t-channel cross sections have been measured at Tevatron by $CDF$ and $DO$ collaborations and the
measurement precision can reach $18\%$ [18]. The measurement precision for the t-channel cross section at the
$8TeV$ $LHC$ reported by $ATLAS$ and $CMS$ is about $15\%$ [19]. It will be enhanced in coming years. For
example, Ref.[34] has shown that the cross section of the t-channel single top production at the $14TeV$ $LHC$
 can be measured with a precision of $5\%$.

\begin{figure}[htb]
\begin{center}
 \epsfig{file=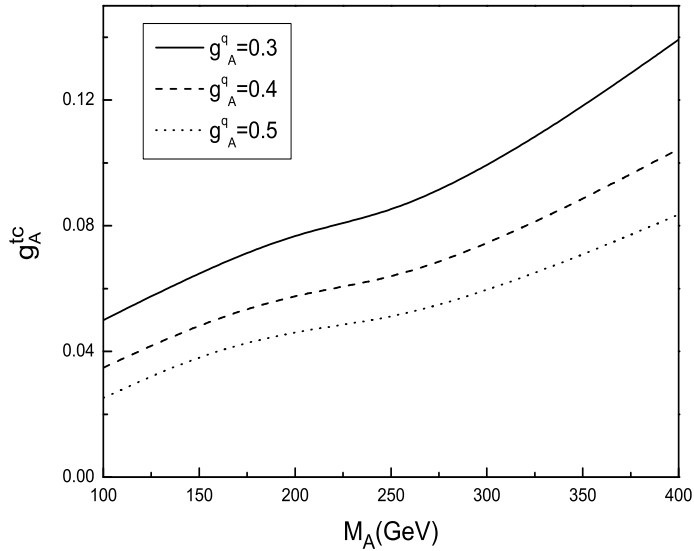,width=300pt,height=260pt}
\vspace{-0.5cm} \caption{In the case of $\delta\sigma^{t}/\sigma_{SM}^{t}=10\%$, the $FC$ coupling $g_{A}^{tq}$ as function of the axigluon \hspace*{1.7cm} mass $M_{A}$ for $g_{A}^{q}=0.3$(solid line), $0.4$(dashed line) and $0.5$(dotted line). } \label{ee}
\end{center}
\end{figure}

From above discussions we can see that the theoretical error of the
$SM$ $ NNLO$ cross section at the $14TeV$ $LHC$ for the s- and t-channel productions could be as large as $5\%$,
the same amount of the expected precision  at the $14TeV$ $LHC$. So if the relative correction of the light axigluon to the single top production
cross section is larger than $10\%$, the $14TeV$ $LHC$ should detect this correction effect. In Fig.4 and Fig.5 we demand that $\delta\sigma^{s}/\sigma_{SM}^{s}=10\%$ and $\delta\sigma^{t}/\sigma_{SM}^{t}=10\%$, where $\sigma_{SM}^{s}$ and $\sigma_{SM}^{t}$ are the $SM$ $ NNLO$
predictions for the s- and t-channel single top production cross sections at the $LHC$ with $\sqrt{s}=14TeV$,
$\delta\sigma^{s}$ and $\delta\sigma^{t}$ are induced by the light axigluon $A$, and plot the $FC$ coupling
$g_{A}^{tq}$ as a function of the mass parameter $M_{A}$ for different values of the flavor conserving $g_{A}^{q}$.
In our numerical calculation, we have taken the central values for $\sigma_{SM}^{s}$ and $\sigma_{SM}^{t}$. From
these figures one can see that the contributions of the light axigluon to the production cross sections of the
processes $pp\rightarrow tb+X$ and $pp\rightarrow tj+X$ increase as the coupling parameters $g_{A}^{tq}$ and
$g_{A}^{q}$ increasing, while decrease as $M_{A}$ increasing. For $100GeV\leq M_{A}\leq400GeV$ and $0.3\leq g_{A}^{q}\leq0.5$,
the values of $FC$ coupling $g_{A}^{tq}$ are in the ranges of $0.017\sim0.163$ and $0.024\sim0.139$ for
$\delta\sigma^{s}/\sigma_{SM}^{s}=10\%$ and $\delta\sigma^{t}/\sigma_{SM}^{t}=10\%$, respectively.
We expect that, in near future, the $LHC$ can authenticate this correction effect on single top production
or at least give constraint on the $FC$ coupling $g_{A}^{tq}$.

\vspace{0.5cm} \noindent{\bf 4. The light axigluon and the rare top decays $t\rightarrow c\gamma$ and $cg$ }

\vspace{0.5cm}It is well known that in the $SM$ the rare top decays $t\rightarrow qV$ ($q=u,c$ and $V=\gamma,g,Z$)
mediated by $FCNCs$ are highly $GIM$ suppressed with branching ratios of
$Br(t\rightarrow cV)\sim10^{-14}\sim10^{-12}$ [35], which are far below the detectable level of current
or near future experiments. However, some new physics models can enhance these branching ratios significantly [36].
So rare top decays offer an opportunity to test the $SM$ and search for new physics effects. Any positive signal
of rare top decay processes would clearly indicate new physics beyond the $SM$.
\vspace{0.5cm}
\begin{figure}[htb]
\begin{center}
 \epsfig{file=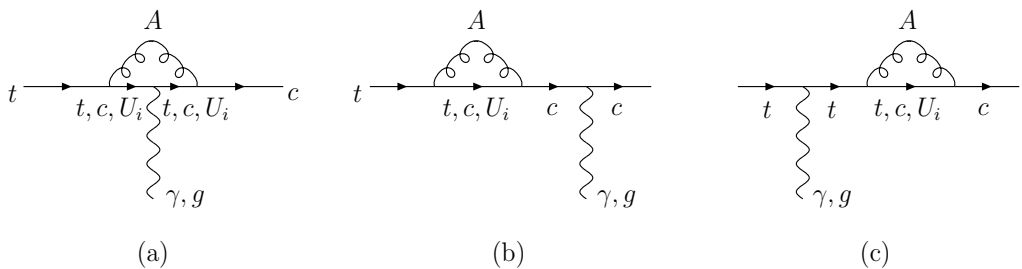,width=380pt,height=100pt}
\vspace{-0.5cm} \caption{Feynman diagrams for the rare top decays $t\rightarrow c\gamma$ and $cg$ coming from the $FC$ \hspace*{1.7cm} coupling $g_{A}^{tc}$, in which $i=1 $ and $2$. } \label{ee}
\end{center}
\end{figure}

On the experimental side, rare top decays are being searched for at Tevatron [37] and $LHC$ [38, 39]. $ATLAS$
collaboration has set upper limit on the branching ratio $Br(t\rightarrow cg)<2.7\times10^{-4}$ at $95\%$ C.L. [39].
The sensitivity of $ATLAS$ to the branching ratio $Br(t\rightarrow c\gamma)$ is expected to be of the order of
$10^{-4}$ [40].

From discussions given in above sections we can see that the light axigluon with $FC$ couplings can contribute rare
top decays. In this section we will calculate the branching ratios $Br(t\rightarrow c\gamma)$ and $Br(t\rightarrow cg)$
induced by the light axigluon. The relevant Feynman diagrams are shown in Fig.6. In this section, we also assume that the contributions of the third generation new quarks to the rare top decays   $t\rightarrow c\gamma$ and $t\rightarrow cg$ decouple. Compared to the $FC$ couplings  of the light axigluon $A$ to  the new quarks and the $SM$ quarks,  the $FC$ couplings  of the scalar $\phi$ to  the new quarks and the $SM$ quarks arise at higher order, their $FC$ effects are much smaller than those induced by the axigluon $A$. Thus, in this section, we neglect the contributions of the scalar $\phi$ to the rare top decays  $t\rightarrow c\gamma$ and $t\rightarrow cg$ as done for $Z\rightarrow bs$ in section 2.

Considering electromagnetic gauge invariance, the amplitude of the rare decay $t\rightarrow c\gamma$ can be general
written as
\begin{eqnarray}
{M(t\rightarrow c\gamma)=i\overline{u}(P_{c})\sigma^{\mu\nu}q_{\nu}(A_{\gamma}+B_{\gamma}\gamma_{5})u(P_{t})\varepsilon_{\mu}^{\ast}(q)},
\end{eqnarray}
\begin{figure}[htb]
\begin{center}
 \epsfig{file=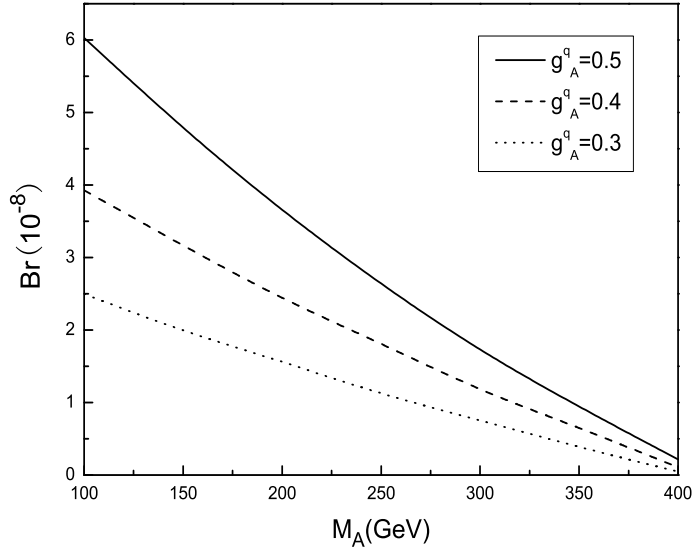,width=300pt,height=260pt}
\vspace{-0.5cm} \caption{The branching ratio $Br(t\rightarrow c\gamma)$ as a function of the axigluon mass $M_{A}$ for
\hspace*{1.8cm} three values of the flavor conserving coupling $g_{A}^{q}$. } \label{ee}
\end{center}
\end{figure}
where $q=P_{t}-P_{c}$ is the photon momentum and $\varepsilon$ is its polarization vector, in which $P_{t}$
and $P_{c}$ represent the momenta of top and charm quarks, respectively. A similar structure is valid
for $t\rightarrow cg$ with form factors $A_{g}$ and $B_{g}$. For the light axigluon $A$ with zero vector
couplings to the $SM$ and new quarks i.e. $g_{V}^{tq}\approx0$, $g_{V}^{Q_{H}q}\approx0$ and $g_{V}^{q}\approx0$ [10, 12], there are $A_{\gamma}\neq0$, $A_{g}\neq0$
and $B_{\gamma}=0$, $B_{g}=0$. Recently, Ref.[41] has calculated the contributions of color-singlet
gauge bosons predicted by the 331 models to the rare top decay $t\rightarrow c\gamma$ and give the
explicit expressions for the relevant form factors. In this paper we will use LoopTools [42] to
obtain our numerical results.

Using Eq.(11), the partial widths of $t\rightarrow c\gamma$ and $t\rightarrow cg$ contributed
by the light axigluon can be written as
\begin{eqnarray}
{\Gamma(t\rightarrow c\gamma)=\frac{m_{t}^{3}}{8\pi}(1-\frac{m_{c}^{2}}{m_{t}^{2}})^{3}|A_{\gamma}|^{2}},
\end{eqnarray}
\begin{eqnarray}
{\Gamma(t\rightarrow cg)=\frac{C_{F}m_{t}^{3}}{8\pi}(1-\frac{m_{c}^{2}}{m_{t}^{2}})^{3}|A_{g}|^{2}},
\end{eqnarray}
where $C_{F}  = 4/3 $ is a color factor.

\begin{figure}[htb]
\begin{center}
 \epsfig{file=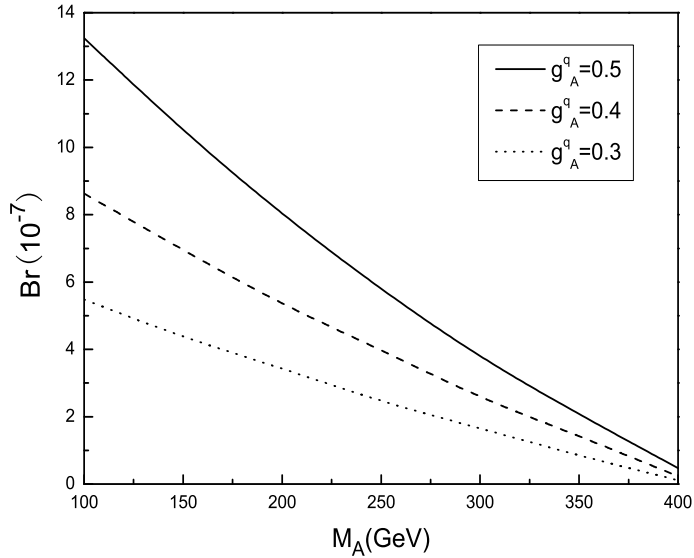,width=300pt,height=260pt}
\vspace{-0.5cm} \caption{The branching ratio $Br(t\rightarrow cg)$ as a function of the axigluon mass $M_{A}$ for
\hspace*{1.8cm} three values of the flavor conserving coupling $g_{A}^{q}$.} \label{ee}
\end{center}
\end{figure}

To obtain numerical results, we have assumed that the top total decay width is dominated by the
decay $t\rightarrow Wb$. The $FC$ coupling $g_{A}^{tc}$ is determined by the parameters $g_{A}^{q}$
and $M_{A}$ via the relation $\delta\sigma^{t}/\sigma_{SM}^{t}=10\%$. For calculation the contributions of the first and second generation new quarks, we take
the case II: $\varepsilon_{Hd} = I, ~~ \varepsilon_{Hu}= V_{CKM}$ and assume $M_{H}=0.2M_{A}$.
In Fig.7 and Fig.8 we plot the branching ratios $Br(t\rightarrow c\gamma)$ and $Br(t\rightarrow cg)$ as functions of the axigluon
mass $M_{A}$ for three values of the flavor conserving coupling $g_{A}^{q}$. One can see from these figures that the light axigluon
$A$ can indeed enhance the branching ratios $Br(t\rightarrow c\gamma)$ and $Br(t\rightarrow cg)$. For $0.3\leq g_{A}^{q}\leq0.5$ and $100GeV\leq M_{A}\leq400GeV$, the values of $Br(t\rightarrow c\gamma)$ and $Br(t\rightarrow cg)$ are in the ranges of
$4.8\times10^{-9}\sim 5.9\times10^{-8}$ and $1.1\times10^{-8}\sim 1.3\times10^{-6}$, respectively. Replacing the $FC$ couplings $g_{A}^{tc}$ and $ g_{A}^{U_{i}c}$ by $g_{A}^{tu}$ and $ g_{A}^{U_{i}u}$, we can easily calculate the contributions of the light axigluon $A$ to the rare top decays $t\rightarrow u\gamma$ and $ug$.

\vspace{0.5cm} \noindent{\bf 5. Conclusions }

\vspace{0.5cm}The light axigluon $A$ with a mass $M_{A}$ in the
range from $100GeV$ to $400GeV$ predicted by the light axigluon model [10] can explain the $t\overline{t}$ $FBA$ and satisfy
the constraints from the $ATLAS$ and $CMS$ data, as long as its decay width
is large and its couplings to the $SM$ quarks are relatively small. In order to get suppressed couplings of the light axigluon $A$ to the $SM$ quarks, the new quarks and the $SM$ quarks should have mixing, which can induce the $FC$ couplings to  the new quarks and the $SM$ quarks. Furthermore, to fulfill the broad width of the axigluon, the new quarks, at least the first and second generation new quarks, are lighter than the light axigluon. In this paper, we assume the flavor
conserving axigluon couplings are universal and pure axial vector-like, and investigate
some $FC$ phenomena mediated by the light axigluon.

The contributions of the light axigluon model to the $FC$ decays  $Z\rightarrow \overline{b}s,b\overline{s}$ and  $t\rightarrow c\gamma, cg$ mainly come from the $FC$ quark- quark- axigluon coupling $g_{A}^{qq'}$ and  the $FC$  quark- new quark- axigluon coupling $g_{A}^{qQ_{H}}$. Considering the constraints of meson  mixing on the $FC$ coupling $g_{A}^{qq'}$ and assuming that both
$\varepsilon_{Hu}$ and $\varepsilon_{Hd}$ are nearly equal to the identity matrices and satisfy the relation $\varepsilon^{+}_{Hu}\varepsilon_{Hd}= V_{CKM}$ to give the value of  $g_{A}^{qQ_{H}}$, we calculate
the branching ratios $Br(Z\rightarrow \overline{b}s+b\overline{s})$, $Br(t\rightarrow c\gamma)$ and $Br(t\rightarrow cg)$ in the context of the light axigluon model.
Our numerical results show that, in most of parameter space,  the value of the branching ratio $Br(Z\rightarrow \overline{b}s+b\overline{s})$
is smaller than $1\times10^{-8}$, which is still below the $SM$ prediction. Compared to the $SM$ predictions, the branching ratios $Br(t\rightarrow c\gamma)$ and $Br(t\rightarrow cg)$ can be significantly enhanced in the light axigluon model, while are still lower than the corresponding current
experimental upper limits.

It is well known that single top production is very sensitive to new physics beyond the $SM$, whose effects
can be assessed by precise measurement of the production cross section. In this paper, we study the correction
effects of the light axigluon $A$ to the s- and t-channel single top productions at the $LHC$. We find that, in
near future, the $LHC$ should observe this correction effect with reasonable values for the $FC$ coupling $g_{A}^{tq}$ or at least give constraint on the $FC$ coupling $g_{A}^{tq}$.
If one demands $\delta\sigma^{s}/\sigma_{SM}^{s}=10\%$ and $\delta\sigma^{t}/\sigma_{SM}^{t}=10\%$, the values of the
$FC$ coupling $g_{A}^{tq}$ should be in the ranges of $0.017\sim0.163$ and $0.024\sim0.139$, respectively.

\section*{Acknowledgments} \hspace{5mm}This work was
supported in part by the National Natural Science Foundation of
China under Grants No. 11275088 and Foundation of Liaoning Educational Committee (No. LT2011015).


\vspace{1.0cm}


\begin{thebibliography}{99}

\bibitem{y1}G. Aad et al. [ATLAS Collaboration], {\em Phys. Lett. B} {\bf716} (2012) 1.


\bibitem{y2}S. Chatrchyan et al. [CMS Collaboration ], {\em Phys. Lett. B} {\bf716} (2012) 30.

\bibitem{y3}C. T. Hill, {\em Phys. Lett. B} {\bf266} (1991) 419; C. T. Hill, {\em Phys. Lett. B} {\bf345} (1995) 483;
            R. Chivukula, A. G. Cohen, and E. H. Simmons, {\em Phys. Lett. B} {\bf380} (1996) 92; E. H. Simmons, {\em Phys. Rev. D} {\bf55} (1997) 1678.

\bibitem{y4}J. C. Pati and A. Salam, {\em Phys. Rev. Lett. } {\bf34} (1975) 613; P. H. Frampton and S. L. Glashow, {\em Phys. Lett. B} {\bf190} (1987) 157; {\em Phys. Rev. Lett. } {\bf 58} (1987) 2168; J. Bagger, C. Schmidt, and S. King, {\em Phys. Rev. D} {\bf37} (1988) 1188; R. S. Chivukula, E. H. Simmons, and C. -P. Yuan, {\em Phys. Rev. D} {\bf82} (2010) 094009; P. H. Frampton, J. Shu, and K. Wang, {\em Phys. Lett. B} {\bf683} (2010) 294.

\bibitem{y5}D. Dicus, C. McMullen and S. Nandi, {\em Phys. Rev. D} {\bf65} (2002) 076007.

\bibitem{y6}E. Farhi and L. Susskind, {\em Phys. Rept. } {\bf 74} (1981) 277; K. Lane and S. Mrenna, {\em Phys. Rev. D} {\bf67} (2003) 115011.

\bibitem{y7}T. Aaltonen et al. [CDF Collaboration], {\em Phys. Rev. Lett.} {\bf101} (2008) 202001; T. Aaltonen et al. [CDF Collaboration], {\em Phys. Rev. D } {\bf83} (2011) 112003; V. M. Abazov et al. [D0 Collaboration], {\em Phys. Rev. Lett.} {\bf100} (2008) 142002.

\bibitem{y8}P. Ferrario and G. Rodrigo, {\em Phys. Rev. D} {\bf78} (2008) 094018; M. Martynov and A. Smirnov, {\em Mod. Phys. Lett. A} {\bf24} (2009) 1897;  P. Ferrario and G. Rodrigo, {\em Phys. Rev. D} {\bf80} (2009) 051701; Q. -H. Cao, D. McKeen, J. L. Rosner, G. Shaughnessy, and C. E. Wagner, {\em Phys. Rev. D} {\bf81} (2010) 114004; R. Chivukula, E. H. Simmons, and C. -P. Yuan, {\em Phys. Rev. D} {\bf82} (2010) 094009; Y. Bai, J. L. Hewett, J. Kaplan, and T. G. Rizzo, {\em JHEP }  {\bf1103} (2011) 003;  A. R. Zerwekh, {\em Phys. Lett. B} {\bf704} (2011) 62; M. I. Gresham, I. -W. Kim, and K. M. Zurek, {\em Phys. Rev. D} {\bf83} (2011) 114027; A. Djouadi, G. Moreau, and F. Richard, {\em Phys. Lett. B} {\bf701} (2011) 458; Junjie Cao, Lei Wu, Jin Min Yang, {\em Phys. Rev. D} {\bf83} (2011) 034024; E. Alvarez, L. Da Rold, J. I. S. Vietto, and A. Szynkman, {\em JHEP } {\bf1109} (2011) 007; J. Aguilar-Saavedra and M. Perez-Victoria, {\em Phys. Lett. B} {\bf705}(2011) 228; H. Wang, Y. -K. Wang, B. Xiao, and S. -H. Zhu, {\em Phys. Rev. D} {\bf84} (2011) 094019; G. Z. Krnjaic, {\em Phys. Rev. D} {\bf85} (2012)014030.

\bibitem{y9}J. Drobnak, J. F. Kamenik and J. Zupan, {\em Phys. Rev. D} {\bf86} (2012) 054022; M. Cvetic, J. Halverson and P. Langacker, {\em JHEP } {\bf1211} (2012) 064; B. D¨ªaz, A. R. Zerwekh, Int. J. Mod. {\em Phys. A} {\bf28} (2013) 1350133.

\bibitem{y10}G. M. Tavares and M. Schmaltz, {\em Phys. Rev. D} {\bf84} (2011) 054008;  C. Gross,  G. M. Tavares, M. Schmaltz,  and C. Spethmann, {\em Phys. Rev. D} {\bf87} (2013) 014004.

\bibitem{y11} U. Haisch and S. Westhoff, {\em JHEP} {\bf1108} (2011) 088.

\bibitem{y12} M. Gresham, J. Shelton, K. M. Zurek, {\em JHEP } {\bf1303} (2013) 008.

\bibitem{y13} S. Dutta, A. Goyal, and M. Kumar,  {\em Phys. Rev. D} {\bf87} (2013) 094016.

\bibitem{y13} G. Aad et al. [ATLAS Collaboration], {\em New J. Phys.} {\bf13} (2011) 053044; {\em Eur. Phys. J. C} {\bf71} (2011) 1828; {\em Phys. Lett. B} {\bf708} (2012) 37; ATLAS Collaboration, ATLAS-CONF-2012-096 (2012); ATLAS-CONF-2012-110 (2012).

\bibitem{y15} V. Khachatryan et al. [CMS Collaboration], {\em Phys. Rev. Lett. } {\bf105} (2010) 211801; S. Chatrchyan et al. [CMS Collaboration], {\em Phys. Lett. B} {\bf704} (2011) 123; {\em Phys. Lett. B} {\bf717}  (2012) 129; {\em JHEP } {\bf1301} (2013) 013; {\em Phys. Rev. D} {\bf87} (2013) 114015.

\bibitem{y16}  W. Bernreuther, {\em J. Phys. G} {\bf35} (2008) 083001; J. R. Incandela,  A. Quadt, W. Wagner, D. Wicke, {\em Prog. Part. Nucl. Phys. } {\bf63} (2009) 239; F. P. Schilling, {\em Int. J. Mod. Phys. A} {\bf27} (2012) 1230016.
\bibitem{y17} T. M. P. Tait and C. P. Yuan, {\em Phys. Rev. D} {\bf63} (2000) 014018; E. Boos, L. Dudko, {\em Int. J. Mod. Phys. A} {\bf27} (2012) 1230026; P. Falgari, {\em J. Phys. Conf. Ser. } {\bf452} (2013) 012016.
\bibitem{y18}  V. M. Abazov et al. [D0 Collaboration], {\em Phys. Rev. Lett. } {\bf103} (2009) 092001; T. Aaltonen et al. [CDF Collaboration], {\em Phys. Rev. Lett. } {\bf103} (2009) 092002; V. M. Abazov et al. [D0 Collaboration], {\em Phys. Lett. B} {\bf705} (2011) 313; {\em Phys. Rev. D} {\bf84} (2011) 112001; {\em Phys. Lett. B} {\bf726} (2013) 656.

\bibitem{y19} G. Aad et al. [ATLAS Collaboration], {\em Phys. Lett. B} {\bf717} (2012) 330; ATLAS-CONF-2012-132 (2013); S. Chatrchyan et al. [CMS Collaboration], {\em Phys. Rev. Lett. } {\bf107} (2011) 091802; {\em JHEP } {\bf1212} (2012) 035; CMS-PAS-TOP-12-011 (2013).

\bibitem{y20} G. Aad et al. [ATLAS Collaboration], {\em Phys. Lett. B} {\bf716} (2012) 142;  S. Chatrchyan et al. [CMS Collaboration], {\em Phys. Rev. Lett.} {\bf110} (2013) 022003.


\bibitem{y21} S. Ipek, {\em Phys. Rev. D} {\bf87} (2013) 116010.

\bibitem{y22} B. A. Dobrescu, K. Kong, and R. Mahbubani, {\em Phys. Lett. B} {\bf670} (2008) 119; M. Cvetic, J. Halverson, and P. Langacker, {\em JHEP} {\bf1211} (2012) 064.
\bibitem{y23} J. Hubisz, S. J. Lee and G. Paz, {\em JHEP} {\bf0606} (2006) 041.

\bibitem{y24} B. Holdom, {\em Phys. Lett. B} {\bf351} (1995) 279.

\bibitem{y25} C. T. Hill and X. M. Zhang, {\em Phys. Rev. D} {\bf51} (1995) 3563.

\bibitem{y26} M. Clements et al., {\em Phys. Rev. D.} {\bf27} (1983) 570; V. Ganapathy et al., {\em Phys. Rev. D.} {\bf27} (1983) 579; W. S. Hou, N. G. Deshpande, G. Eilam and A. Soni, {\em Phys. Rev. Lett. } {\bf57} (1986) 1406; J. Barnabeu, M. B. Gavela and A. Santamaria, {\em Phys. Rev. Lett. } {\bf57} (1986).

\bibitem{y27} J. A. Aguilar-Saavedra et al., hep-ph/0106315; H. Baer et al., arXiv:1306.6352 [hep-ph].

\bibitem{y28} M. Duncan, {\em Phys. Rev. D} {\bf31} (1985) 1139; F. Gabbiani, J. H. Kim and A. Massiero, {\em Phys. Lett. B} {\bf214} (1988) 398; C. Bush, {\em Nucl. Phys. B} {\bf319} (1989) 15; B. Mukhopadhyaya  and A. Raychoudhuri, {\em Phys. Rev. D} {\bf39} (1989) 280; W. S. Hou and R. G. Stuart, {\em Phys. Lett. B} {\bf226} (1989) 122; B. Grzadkowski, J. F. Gunion and P. Krawczyk, {\em Phys. Lett. B} {\bf268} (1991) 106; Xue-Lei Wang, Gong-Ru Lu, Zhen-Jun Xiao, {\em Phys. Rev. D} {\bf51} (1995) 4992; D. Atwood, L. Reina and A. Soni, {\em Phys. Rev. D} {\bf55} (1997) 3156; D. Atwood, S. Bar-Shalom, G. Eilam and A. Soni, {\em Phys. Rev. D} {\bf66} (2002) 093005; Chong-Xing Yue, Hong Li, Hong-Jie Zong, {\em Nucl. Phys. B} {\bf650} (2003) 290;  R. Mohanta, {\em Phys. Rev. D} {\bf71} (2005) 114013; Xiao-Fang Han, Lei Wang, Jin Min Yang, {\em Phys. Rev. D} {\bf78} (2008) 075017.

\bibitem{y29} J. Beringer et al. [Particle Data Group], {\em Phys. Rev. D} {\bf86} (2012) 010001.

\bibitem{y30} D. Karabacak, S. Nandi, S. K. Rai, {\em Phys. Rev. D} {\bf85} (2012) 075011


\bibitem{y31} J. Pumplin et al, {\em JHEP } {\bf0207} (2002) 012.

\bibitem{y32} S. S. D. Willenbrock and D. A. Dicus, {\em Phys. Rev. D} {\bf34} (1986) 155.

\bibitem{y33} N. Kidonakis, {\em Phys. Rev. D} {\bf74} (2006) 114012; {\em Phys. Rev. D} {\bf75} (2007) 071501(R); {\em Phys. Rev. D} {\bf83} (2011) 091503(R); arXiv:1205.3453 [hep-ph]; arXiv:1210.7813 [hep-ph]; arXiv:1212.2844 [hep-ph].

\bibitem{y34} B. Schoenrock, E. Drueke, B. A. Gonzalez, R. Schwienhorst,  arXiv:1308.6307 [hep-ex].

\bibitem{y35} G. Eilam, J. L. Hewett and A. Soni, {\em Phys. Rev. D} {\bf44} (1991) 1473 [Erratum-ibid. D 59, 039901 (1999)]; J. A. Aguilar-Saavedra, {\em Acta Phys. Polon. B} {\bf35} (2004) 2695.

\bibitem{y36} For example, see: F. Larios , R. Martinez, M. A. Perez, {\em Int. J. Mod. Phys. A} {\bf21} (2006) 3473; J. M. Yang, {\em Int. J. Mod. Phys. A} {\bf23} (2008) 3343; J. Drobnak, arXiv:1210.5051 [hep-ph].

\bibitem{y37} T. Aaltonen et al. [CDF Collaboration], {\em Phys. Rev. Lett.} {\bf101} (2008) 192002; V. M. Abazov et al. [D0 Collaboration], {\em Phys. Lett. B} {\bf701} (2011) 313.


\bibitem{y38} G. Aad et al. [ATLAS Collaboration], {\em JHEP } {\bf1209} (2012) 139; S. Chatrchyan et al. [CMS Collaboration],  {\em Phys. Lett. B} {\bf718} (2013) 1252.

\bibitem{y39} G. Aad et al. [ATLAS Collaboration], {\em Phys. Lett. B} {\bf712} (2012) 351.

\bibitem{y40} J. Carvalho et al. [ATLAS Collaboration], {\em Eur. Phys. J. C} {\bf52} (2007) 999.

\bibitem{y41} I. Cortes-Maldonado, G. Hernandez-Tome,  and G. Tavares-Velasco, {\em Phys. Rev. D} {\bf88} (2013) 14011.

\bibitem{y42} T. Hahn, M. Perez-Victoria, {\em Comput. Phys. Commun.} {\bf118} (1999) 153; T. Hahn, {\em Nucl. Phys. Proc. Suppl.} {\bf135} (2004) 333.

\end{thebibliography}
\end{document}